\documentclass[reprint,showpacs,amsmath,amssymb,aps,pra,longbibliography,floatfix,superscriptaddress]{revtex4-1}

\usepackage[utf8]{inputenc}	
\usepackage[T1]{fontenc}	
\usepackage[]{graphicx}		
\usepackage{color}
\usepackage[english]{babel}

\usepackage{url}

\usepackage{microtype}
\usepackage{braket}
\usepackage{esvect}
\usepackage[export]{adjustbox}

\usepackage{amsmath,amsfonts,amssymb}
\usepackage{mathtools}

\usepackage[position=top,caption=false]{subfig}

\let\originaleqref\eqref
\renewcommand{\eqref}{Eq.~\originaleqref}

\newcommand{\fref}[1]{Fig.~\ref{#1}}

\graphicspath{
{figures/}
}

\begin{document}
\title{The quantum regression theorem for out-of-time-ordered correlation functions}

\author{Philip Daniel Blocher}
\email{pblocher@phys.au.dk}
\affiliation{Department of Physics and Astronomy, Aarhus University, Ny Munkegade 120, DK-8000 Aarhus C, Denmark}

\author{Klaus M{\o}lmer}
\affiliation{Department of Physics and Astronomy, Aarhus University, Ny Munkegade 120, DK-8000 Aarhus C, Denmark}

\date{\today}
\bigskip

\begin{abstract}

We derive an extension of the quantum regression theorem to calculate out-of-time-order correlation functions in Markovian open quantum systems. While so far mostly being applied in the analysis of many-body physics, we demonstrate that out-of-time-order correlation functions appear naturally in optical detection schemes with interferometric delay lines, and we apply our extended quantum regression theorem to calculate the non-trivial photon counting fluctuations in split and recombined signals from a quantum light source.
\end{abstract}

\maketitle
\noindent

\section{Introduction}
Spatial and temporal correlation functions find application in analyses within e.g. signal processing, financial studies, and a variety of observational studies in natural sciences. Experiments performed by Hanbury-Brown and Twiss in 1956 \cite{HanburyBrownTwiss1956} inspired a general analysis of how the temporal correlations in photodetection signals depend on the dynamics of the emitters, and it was realized that signal correlations may witness their genuine quantum behavior. Anti-bunching and sub-Poissonian counting statistics are thus incompatible with a classical description of the field. A quantum treatment of the photodetection process, taking into account how continuous monitoring of the quantum system causes measurement back action and thus quenches the system at every infinitesimal time step, reveals the source of correlations in the measurement signal \cite{Carmichael1993}. The correlation functions present the average behavior of the correlations and form a bridge between stochastic single-shot trajectories and deterministic methods \cite{Xu2015}.

In Glauber's photodetection theory \cite{Glauber1963,Loudon2000}, light detection is described by the annihilation of a photon and excitation of an electron within the detector, with a subsequent amplification of the electron signal to a classical current. A sequence of such successive annihilation events at times $t_1 \leq t_2 \leq \dots \leq t_N$ has a probability given by the normal-ordered correlation function of the field annihilation and creation operators, $\braket{a^\dagger(t_1)\dots a^\dagger(t_N) a(t_N) \dots a(t_1)}$. This particular intuitive structure, with field creation operators to the left and field annihilation operators on the right and with time increasing towards the middle of the expression, emphasizes the evolution of the quantum field as one photon is removed at a time in successive detection events.

In this article we deal with general quantum correlation functions on the form
\begin{equation}
\braket{A(t_1) B( t_2) C( t_3) \dots}, \label{eq:1}
\end{equation}
where $A$, $B$, $C$, \dots, are arbitrary operators of a quantum system and where the time arguments are not ordered. Recently, there has been a growing interest in the so-called out-of-time-order correlation functions (OTOCs) of a particular form
\begin{equation}
K(t) = \braket{A(t) B(0) C(t) D(0)}, \label{eq:chaosOTOC}
\end{equation}
where $A$, $B$, $C$, and $D$ are Hermitian operators. $K(t)$ reveals how the Hamiltonian of a system propagates disturbances and correlations, and it was first applied to describe the failure of the electron momentum operator to commute at different times in superconducting systems with disorder \cite{Larkin1969}. $K(t)$ finds use today in the diagnosis of scrambling and spreading of quantum information in many-body quantum systems \cite{Garttner2017,Garttner2018,Syzranov2018}, i.e., as a measure of how fast local perturbations become inaccessible to local probes or as an entanglement witness. Within the field of quantum chaos, expressions like \eqref{eq:chaosOTOC} with unitary rather than Hermitian operators quantify the sensitivity of a system's evolution to its initial conditions \cite{YungerHalpern2017}.

\begin{figure}[htp]
\centering
\includegraphics[width = 1\linewidth]{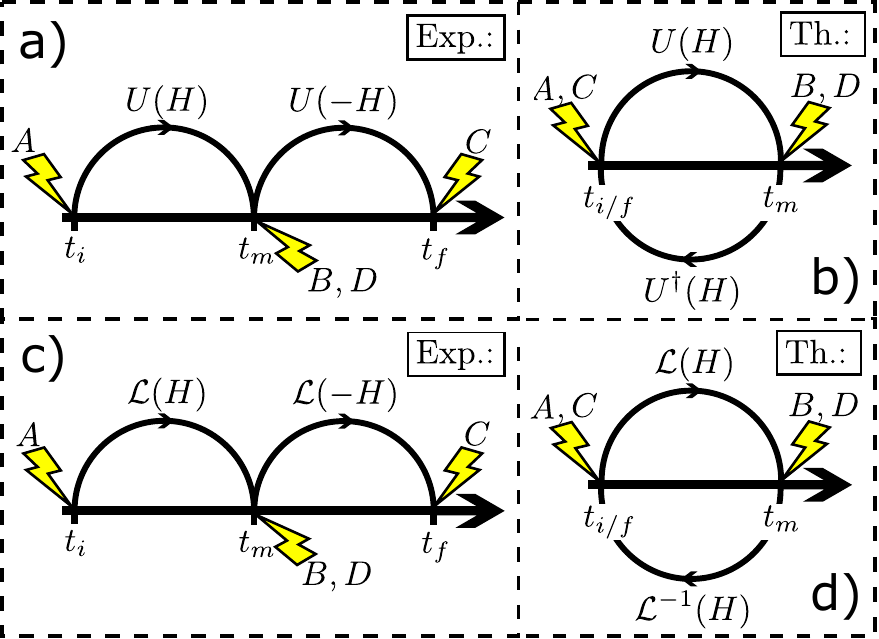}
\caption{Illustration of the correlation function \eqref{eq:chaosOTOC} for experiments (left) and theory (right) in the case of \textbf{a-b)} closed quantum systems, and \textbf{c-d)} open quantum systems. The lightning bolts indicate the evaluation of operators. For closed quantum systems unitary time evolution yields $U^\dagger(H) = U(-H)$. For open quantum systems, interactions with the environment -- e.g. decay -- break time reversibility, and we are unable to engineer the inverse time Liouvillian evolution operator $\mathcal{L}^{-1}$ by merely changing the sign of $H$. A different analysis and experimental investigation of \eqref{eq:chaosOTOC} is therefore needed.}
\label{fig:introIllustration}
\end{figure}

While the theoretical expression \eqref{eq:chaosOTOC} makes sense in the Heisenberg picture of quantum mechanics, it does not provide a recipe for the measurement of the observables. In particular, it does not explain how one can measure $D$ at time $0$, then $C$ at time $t$, and then go back in time to measure $B$ at time $0$, before finally measuring $A$ at the final time $t$. An isolated quantum system evolves under a unitary operation, $U=\exp(-iHt/\hbar)$, and reversal of the sign of $H$ yields the same effect as reversing the sign of the time argument. OTOCs on the form \eqref{eq:chaosOTOC} have thus been measured in Ising model quantum simulators \cite{Garttner2017}, implementing time reversal by merely changing the sign of the Hamiltonian $H \rightarrow -H$. This is familiar to dynamical decoupling strategies, and in \cite{Garttner2017} the OTOC \eqref{eq:chaosOTOC} appears as a fidelity measurement, implementing a many-body Loschmidt echo. Theoretically the unitary time evolution allows \eqref{eq:chaosOTOC} to be expressed as a process of moving forward and backward in time to apply operators at their appropriate times. In \fref{fig:introIllustration}, panels \textbf{a} and \textbf{b} illustrate the experimental and theoretical implementation of \eqref{eq:chaosOTOC} in the case of a closed quantum system subject to unitary time evolution.

While closed quantum systems permit studies of OTOCs by engineering of the Hamiltonian, we want to extend the analysis to open quantum systems interacting with environment degrees of freedom. This provides more realistic models especially for experimental work, where system-environment interactions lead to decay and decoherence of the quantum system, and it connects to the more general theory of quantum measurements using ancillary degrees of freedom, such as the monitoring of a two-level atom by detection of its emitted fluorescence. As we cannot feasibly reverse the system-environment interaction, reversal of the time evolution in experiments is unavailable. We illustrate this point in \fref{fig:introIllustration} panels \textbf{c} and \textbf{d}, where application of the experimental Loschmidt echo procedure is no longer equivalent to the theoretical backward time evolution -- the time reversed Liouvillian $\mathcal{L}^{-1}(H)$ differs from the forward evolution under $\mathcal{L}(-H)$.

The correlation functions are still well-defined in the Heisenberg picture, and some attention has recently been drawn to the lack of unitary time evolution for systems weakly coupled to dissipative environments~\cite{Syzranov2018}. Under the Born-Markov assumption for the system-bath interaction, the so-called quantum regression theorem (QRT) relates the time evolution of normal-ordered correlation functions to the time evolution of the system's density matrix. In this article we shall derive a generalization of the QRT to determine OTOCs for Markovian open quantum systems.

We also comment on the OTOCs being seemingly at variance with the inherently normal-ordered physical processes of measuring photonic signals. In fact, while the detector signal relates to the annihilation of photons in normal order, the light emitted from an emitter may have been subjected to different propagation delays in interferometer setups, as e.g. in the works by Schrama \textit{et al.} \cite{Schrama1992} and Bali \textit{et al.} \cite{Bali1993} where OTOCs appear. In both of these articles, rather than making use of a unified framework like the QRT, special theory was employed to evaluate the OTOCs. In this article we shall demonstrate the use of our generalized QRT in the analysis of an interferome-tric optical setup, in which several propagation paths of different propagation times exist between the source and the detector, hence allowing the order of photon emission to be interchanged with respect to the order of photon detection.

The structure of this article is as follows: In section~\ref{section:QRTextension} we extend the QRT to multi-time correlation functions of arbitrary temporal ordering, and we emphasise the role of noise contributions in the time evolution. In section~\ref{section:application} we explore how OTOCs occur in quantum optics by considering an interferometer that induces a relative time delay, and we apply our extended QRT to determine intensity correlation of the fluorescence from a two-level atom after transmission through the interferometer. Finally, in section~\ref{section:discussion} we provide an outlook on the interplay between quantum optics studies and recent research in OTOCs.

\section{Extension of the quantum regression theorem}\label{section:QRTextension}
\subsection{Derivation}\label{subsection:QRTderivation}
We consider a principal quantum system $S$ with Hamiltonian $H_S$ coupled by the interaction Hamiltonian $H_{SE}$ to a broad band environment with Hamiltonian $H_E$. The system and its environment may in principle be considered as a single closed quantum system described by a density matrix $\rho_{SE}$, however by assuming the validity of the Born-Markov approximation and tracing out the environment degrees of freedom, the quantum system may be described by a density matrix $\rho = \text{Tr}_E(\rho_{SE})$ obeying a linear master equation
\begin{equation}
\frac{\mathrm{d}}{\mathrm{d}t} \rho_{ij}(t) = \sum_{i^\prime j^\prime} M_{ij, i^\prime j^\prime} \, \rho_{i^\prime j^\prime}(t), \label{eq:densityMatrixMasterEq}
\end{equation}
where $\rho_{ij}(t) = \braket{(\ket{j}\bra{i})(t)}$. For a system of dimension $N$, $M$ is a $N^2 \times N^2$ matrix.
Any linear system operator $A$ is specified by its action on all basis states, and given an orthonormal basis $\{\ket{i}\}$, we may expand $A(t) = \sum_{ij} A_{ji} (\ket{j}\bra{i})(t)$. This in turn implies the well known $\braket{A(t)} = \sum_{ij} \rho_{ij}(t) A_{ji}=$Tr$(\rho A)$.

To address the temporal operator correlation functions, we note that the master equation is equivalent to a set of coupled operator equations of motion in the Heisenberg picture
\begin{equation}
\frac{\mathrm{d}}{\mathrm{d}t}(\ket{j}\bra{i})(t) = \sum_{i^\prime j^\prime} M_{ij, i^\prime j^\prime} (\ket{j^\prime}\bra{i^\prime})(t) + F_{ij}(t). \label{eq:dyadicEoM}
\end{equation}
Here the noise operators $F_{ij}$ have environment operator character and they appear due to the interaction to first order between the system and environment. These operator terms are uncorrelated with the previous evolution of the system and have zero mean. They therefore do not contribute in \eqref{eq:densityMatrixMasterEq}, but are necessary to ensure a consistent formulation and e.g. preserve operator identities such as commutator relations.

Let us recall how two-time averages on the form $\braket{A(t) B(t+\tau)}$ are calculated for $\tau \geq 0$ using the QRT \cite{Cohen-Tannoudji1992}. We expand the operator $B(t+\tau)$ on the set of dyadic products, $B(t+\tau) = \sum_{ij} B_{ji} (\ket{j}\bra{i})(t+\tau)$, and write
\begin{align}
\braket{A(t) B(t+\tau)} =& \text{Tr}(\rho_A(t+\tau) B) \nonumber\\
=& \sum_{ij} \rho_{A,ij}(t+\tau) B_{ji},
\end{align}
where we have defined the matrix
\begin{equation}
\rho_{A,ij}(t+\tau) = \braket{A(t) (\ket{j}\bra{i})(t+\tau)}. \label{eq:twoTimeCorrelationFunctionObject}
\end{equation}
The equations of motion for the matrix elements in  \eqref{eq:twoTimeCorrelationFunctionObject} follow from \eqref{eq:densityMatrixMasterEq} as
\begin{align}
\frac{\mathrm{d}}{\mathrm{d}\tau} \rho_{A,ij}(t+\tau) =& \frac{\mathrm{d}}{\mathrm{d}\tau} \braket{A(t) (\ket{j}\bra{i})(t+\tau)} \nonumber\\
=& \sum_{i^\prime j^\prime} M_{ij, i^\prime j^\prime} \braket{A(t)(\ket{j^\prime}\bra{i^\prime})(t+\tau)} \nonumber\\
& + \braket{A(t) F_{ij}(t+\tau)}.
\end{align}
For $\tau \geq 0$ we have $\braket{A(t) F_{ij}(t+\tau)} = 0$, as $F_{ij}$ is a noise operator with vanishing mean and uncorrelated with the system operator $A$ at previous times. We thus arrive at the familiar formulation of the QRT
\begin{equation}
\frac{\mathrm{d}}{\mathrm{d}\tau} \rho_{A,ij}(t+\tau) = \sum_{i^\prime j^\prime} M_{ij, i^\prime j^\prime} \, \rho_{A,i^\prime j^\prime}(t+\tau), \label{eq:twoTimeCorrelationFunctionObjectEoM}
\end{equation}
namely that the time evolution of the correlation function $\braket{A(t)B(t+\tau)}$ is governed by a matrix that follows from the exact same set of equations as the time evolution of the density matrix (\eqref{eq:densityMatrixMasterEq}) \cite{Cohen-Tannoudji1992,Gardiner2000}.

In order to explore OTOCs for Markovian open quantum systems we require a recipe for calculating multi-time correlation functions on the form \eqref{eq:1} with arbitrary ordering of the time arguments and no assumptions about the nature of the system operators. For the sake of illustration we here consider the case of the out-of-time-ordered correlation function $\braket{A(t+\tau) B(t) C(t+\tau) D(t)}$, $\tau \geq 0$. At time $t+\tau$ we expand the operators $A$ and $C$ on dyadic products and define
\begin{align}
\rho_{ij,B,mn,D}&(t+\tau)\label{eq:fourTimeCorrelationFunctionObject}\\
 =& \braket{(\ket{j}\bra{i})(t+\tau) B(t) (\ket{n}\bra{m})(t+\tau) D(t)} \nonumber
\end{align}
in order that we may write
\begin{align}
\langle A(t+\tau) B(t)& C(t+\tau) D(t) \rangle \nonumber\\
=& \sum_{ij,mn} A_{ji} C_{nm}\, \rho_{ij,B,mn,D}(t+\tau). \label{eq:fourTimeCorrelationFunction}
\end{align}

In \eqref{eq:fourTimeCorrelationFunctionObject} we use a notation such that capital letters denote operators evaluated at the previous time $t$, while the pairs of lowercase letters denote the indices of the constituent dyadic products. The evaluation of \eqref{eq:fourTimeCorrelationFunction} requires first the calculation of the object $\rho_{ij,B,mn,D}(t)$ at the equal time $t$ of all operators -- i.e. a conventional mean value evaluation -- followed by the time propagation of $\rho_{ij,B,mn,D}(t')$ from $t$ to $t+\tau$. Evaluating \eqref{eq:dyadicEoM} to first order in $\mathrm{d}\tau$, one may show that the object \eqref{eq:fourTimeCorrelationFunctionObject} obeys the equation of motion
\begin{align}
\frac{\mathrm{d}}{\mathrm{d}\tau} \rho&_{ij,B,mn,D}(t+\tau) \nonumber\\
=& \sum_{i^\prime j^\prime} M_{ij,i^\prime j^\prime}\, \rho_{i^\prime j^\prime,B,mn,D}(t+\tau) \nonumber\\
+&\sum_{m^\prime n^\prime} M_{mn,m^\prime n^\prime}\, \rho_{ij,B,m^\prime n^\prime,D}(t+\tau) \nonumber\\
+& \mathrm{d}\tau\braket{F_{ij}(t+\tau) B(t) F_{mn}(t+\tau) D(t)}, \label{eq:fourtimeCorrelationFunctionObjectEoM}
\end{align}
where the two first terms on the right-hand side are master equation-like -- i.e. what we would naively have expected -- while the last term is a second order noise contribution to the system evolution. The noise operators $F$ have vanishing mean and are uncorrelated with system observables at earlier times, hence terms linear in $F$ have already been discarded in \eqref{eq:fourtimeCorrelationFunctionObjectEoM}. The challenge then lies in determining the contributions from the product of two noise operators $F$ at equal times. As the number of dyadic products in OTOCs grows so will the number of terms in \eqref{eq:fourtimeCorrelationFunctionObjectEoM}, however the structure will remain unchanged: There will be a number of terms with coefficients inherited from the master equation as well as a number of quadratic noise term contributions.

To evaluate the noise terms, let us write the noise operator $F_{ij}$ as the sum over products of system and environment operators:
\begin{equation}
F_{ij}(t) = \sum_k g_{k,ij} Q_{k}(t) R_{k}(t), \label{eq:noiseStructure}
\end{equation}
where $Q_k$ ($R_k$) is a system (environment) operator and $g_{k,ij}$ is a complex number. Then, without any further assumptions on $Q_k$ or $R_k$, we may separate any noise terms like above in the following manner:
\begin{align}
&\braket{\dots F_{ij}(t) \dots F_{mn}(t)\dots} \nonumber\\
&= \sum_{kl} g_{k,ij} g_{l,mn} \braket{\dots Q_{k}(t) R_k(t)\dots Q_{l}(t) R_l(t)\dots} \nonumber\\
&= \sum_{kl} g_{k,ij} g_{l,mn} \underbrace{\braket{\dots Q_{k}(t) \dots Q_{l}(t)\dots}}_\text{system operators}\cdot \nonumber\\
&\hspace{2.4cm} \cdot \underbrace{\braket{\dots R_k(t) \dots R_l(t)\dots}}_\text{reservoir operators}. \label{eq:noiseSeparation}
\end{align}
Here the `\dots' represent other operators present, either system operators, environment operators, or both. In the last equality we have used that system and environment operators commute: \eqref{eq:fourTimeCorrelationFunctionObject} only contains operators that have already been evaluated at prior times. Hence every system operator in \eqref{eq:noiseSeparation} will be evaluated no later than time $t$, ensuring that our noise operators commute with the system operators. It is now clear that whether \eqref{eq:noiseSeparation} yields finite contributions or vanishes will depend on both the expectation value of system operators as well as the expectation value of the environment operators.

\begin{figure}
\centering
\includegraphics[scale=1]{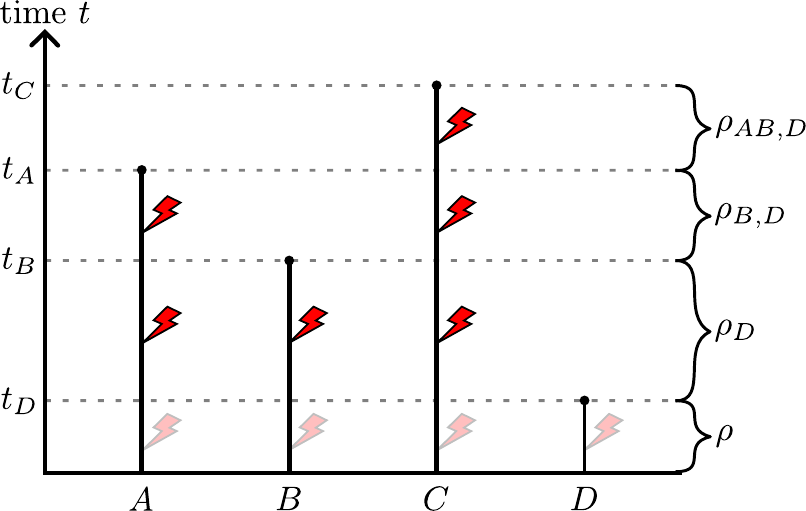}
\caption{Strategy for evaluation of $\braket{A(t_A) B(t_B) C(t_C) D(t_D)}$ for $t_D \leq t_B \leq t_A \leq t_C$. The vertical lines terminate at the times where the particular operators are to be evaluated. On the right hand side of the figure the matrix objects to be time evolved are indicated (see text). Red lightning bolts indicate possible contributions to the time evolution from noise operators; at $t \leq t_D$ the greyed-out lightning bolts indicate that noise terms disappear during the time evolution of $\rho(t)$.}
\label{fig:timeEvolutionStrategy}
\end{figure}

\fref{fig:timeEvolutionStrategy} illustrates the procedure for evaluating a slightly more general correlation function with four different time arguments. In the present case we have $t_A = t_C = t+\tau$, $t_B = t_D = t$, and the procedure thus becomes: Determine the initial state $\rho(t)$, and evaluate the expectation of the operators $B$ and $D$ at time $t$ to create the object $\rho_{ij,B,mn,D}(t)$ using \eqref{eq:fourTimeCorrelationFunctionObject}. Then propagate $\rho_{ij,B,mn,D}(t^\prime)$ in time from $t$ to $t+\tau$ via \eqref{eq:fourtimeCorrelationFunctionObjectEoM}. The correlation function finally follows from \eqref{eq:fourTimeCorrelationFunction}.

The extension of this generalized QRT to general multi-time correlation functions is straight-forward from the procedure described above. In general, correlation functions containing $N$ operators will contain up to $N/2$ dyadic products in the mathematical objects involved in their time evolution.

%

\subsection{Long-time behavior}\label{subsection:QRTregression}
As $\rho(t)$ and $\rho_A(t)$, both required for the calculation of the two-time correlation function $\braket{A(t)B(t+\tau)}$, obey the same linear set of equations, their respective steady states must be proportional. Furthermore, as this set of linear equations preserves trace, we observe the following long-time behavior:
\begin{equation}
\rho_A(t+\tau) \rightarrow \braket{A(t)}\rho^{ss} \text{ for } \tau \rightarrow \infty. \label{eq:twoTimeSteadyState}
\end{equation}
Here $\rho^{ss}$ is the steady state $\rho(t) \rightarrow \rho^{ss}$ for $t \rightarrow \infty$ of the system density matrix, and \eqref{eq:twoTimeSteadyState} implies the decorrelation of two-time averages into products of single-time expectation values at large time separations:
\begin{equation}
\braket{A(t)B(t+\tau)} \rightarrow \braket{A(t)}\braket{B}^{ss} \text{ for } \tau \rightarrow \infty. \label{eq:twoTimeRegression}
\end{equation}

Generally, the noise terms appearing in \eqref{eq:fourtimeCorrelationFunctionObjectEoM} will not all vanish. \eqref{eq:fourtimeCorrelationFunctionObjectEoM} is therefore not guaranteed to be trace preserving and we have no simple expression for its long-time solution. A decorrelation of two-time averages in the case of the OTOC $\braket{A(t+\tau) B(t) C(t+\tau) D(t)}$ only holds if we can disregard contributions from the noise terms in \eqref{eq:fourtimeCorrelationFunctionObjectEoM}. Let us in the following therefore assume these noise terms to be zero at all times; \eqref{eq:fourtimeCorrelationFunctionObjectEoM} then becomes
\begin{align}
\frac{\mathrm{d}}{\mathrm{d}\tau}\rho_{ij,B,mn,D}(t+\tau) =& \sum_{i^\prime j^\prime} M_{ij,i^\prime j^\prime} \; \rho_{i^\prime j^\prime,B,mn,D}(t+\tau) \nonumber\\
+& \sum_{m^\prime n^\prime}  M_{mn,m^\prime n^\prime} \; \rho_{ij,B,m^\prime n^\prime,D}(t+\tau),
\end{align}
which can be shown to preserve trace. The long-time behavior follows as
\begin{equation}
\rho_{ij,B,mn,D}(t + \tau) \rightarrow \braket{B(t)D(t)} \rho_{ij}^{ss} \; \rho_{mn}^{ss} \text{ for } \tau \rightarrow \infty,
\end{equation}
where $\braket{B(t)D(t)} = \text{Tr}[\rho_{B,D}(t)]$, and where $\rho^{ss}$ is the steady state of the system density matrix $\rho(t)$. This implies a decorrelation of the correlation function for large times:
\begin{align}
\langle A(t+\tau)& B(t) C(t+\tau) D(t) \rangle \nonumber\\
&\rightarrow \braket{B(t)D(t)} \braket{A}^{ss} \braket{C}^{ss} \text{ for } \tau \rightarrow \infty,
\end{align}
analogous to the result of \eqref{eq:twoTimeRegression}.

\section{Out-of-time-order correlations in photodetection}\label{section:application}
In this section we demonstrate the application of our extended QRT to the case of a two-level emitter interacting with the electromagnetic radiation field and monitored by photon counting. The emitter may be both coherently driven by a laser field and incoherently pumped by a thermal radiation field. We will show that inserting a Mach-Zender-like interferometer between the emitter and the photodetectors, with one path causing a temporal delay $T$ compared to the other, causes a mixing of the order of detection and emission events. This mixing will be seen to involve the calculation of OTOCS for emitter observables in order to determine the second order correlation function $G^{(2)}$ for detector signals.

\begin{figure}
\centering
\includegraphics[scale=0.9]{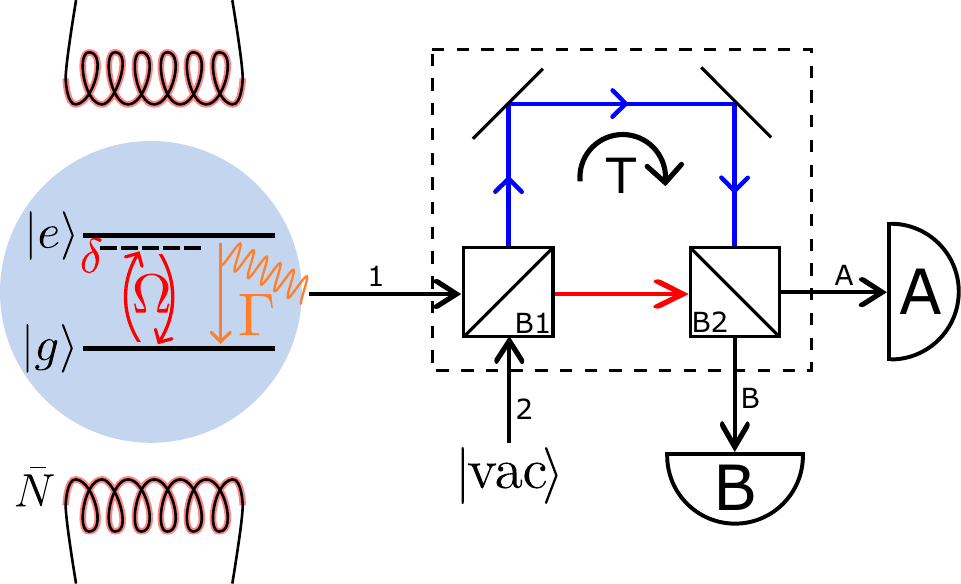}
\caption{A two-level emitter is driven by a monochromatic laser with Rabi frequency $\Omega$ and detuning $\delta$, and decays with a rate $\Gamma$ by interactions with the quantized radiation field. The radiation field may be either in the vacuum state or in a thermal state (indicated by the filaments). In the latter case, the decay rate is $(\bar{N}+1)\Gamma$ and the emitter is incoherently excited at the rate $\bar{N}\Gamma$, augmented by the average thermal photon number $\bar{N}$. Photon counters A and B record the emitted signal at the output of an interferometer which has two inputs 1 (from the emitter) and 2 (vacuum fluctuations). Beamsplitters B1 and B2 split the incoming signals into a short path (red) and a longer path (blue), and recombines the signals at detector A and B). We assume the red path to be of negligible length and the blue path to be a distance $cT$ longer, where $c$ is the speed of light.}
\label{fig:photodetectionSetup}
\end{figure}

\subsection{Setup \& noise terms}
We consider the setup sketched in \fref{fig:photodetectionSetup}: A two-level emitter with ground state $\ket{g}$ and excited state $\ket{e}$ is coupled to both a monochromatic laser field and the quantized electromagnetic radiation field. The radiation field causes incoherent excitation and decay of the emitter, while the laser field serves as a coherent driving field for our emitter. Let $\hbar = 1$ throughout the remainder of this article. In a suitable rotating frame the Hamiltonian for the driven two-level emitter is
\begin{equation}
H_S = -\delta \sigma_+ \sigma_- + \frac{\Omega}{2}(\sigma_+ + \sigma_-),
\end{equation}
where $\delta$ is the laser detuning, $\Omega$ is the Rabi frequency, and $\sigma_- = \ket{g}\bra{e}$ ($\sigma_+ = \sigma_-^\dagger$) is the system's lowering (raising) operator. The radiation field Hamiltonian is $H_E = \sum_k \omega_k a_k^\dagger a_k$, where $a_k$ ($a_k^\dagger$) is the annihilation (creation) operator for a photon in mode $k$, and the coupling between emitter and radiation field is given by the Jaynes-Cummings interaction
\begin{equation}
H_{SE} = \sum_k g_k (\sigma_+ a_k + a_k^\dagger \sigma_-). \label{eq:interactionHamiltonian}
\end{equation}
We assume the environment to be Markovian (memoryless), and in the following we will consider both the case of the radiation field being in the vacuum state $\ket{\text{vac}} \equiv \ket{0}$, and in a broad bandwidth thermal state with a number of quanta per unit bandwidth $\bar{N}$ around the emitter resonance.

The master equation for our system, as well as the noise operators required for the extended QRT, are derived by considering the system and environment operators' equations of motion in the Heisenberg picture \cite{Gardiner2000}. We find that the density matrix obeys the Lindblad master equation
\begin{equation}
\dot{\rho}_S = -i [H_S, \rho_S] + \Gamma (\bar{N}+1)\mathcal{L}[\sigma_-]\rho(t) + \Gamma \bar{N}\mathcal{L}[\sigma_+]\rho(t) ,
\end{equation}
where $\mathcal{L}[c]\rho = c\rho c^\dagger - \frac{1}{2}c^\dagger c \rho - \frac{1}{2}\rho c c^\dagger$ is the Lindblad superoperator, $\Gamma$ is the spontaneous decay/excitation rate, and $\bar{N}$ is the average photon population of the reservoir modes. Combined with the noise terms
\begin{align}
F_{ij}(t) = -& [(\ket{j}\bra{i})(t),\,\sigma_+(t)] a_{in}(t) \nonumber\\
+& [(\ket{j}\bra{i})(t),\,\sigma_-(t)] a_{in}^\dagger(t) \label{eq:noiseTerm}
\end{align}
this yields the equations of motion \eqref{eq:dyadicEoM} for the system dyadic products $(\ket{j}\bra{i})(t)$. In \eqref{eq:noiseTerm} we define the input field $a_{in}(t) = i \sum_k g_k a_k(t_0) \exp[-i \omega_k (t-t_0)]$, where $t_0 \leq t$ is an earlier time at which the environment state and operators are known. The input field correlations
\begin{align}
\braket{a_{in}(t)a_{in}^\dagger(t^\prime)} =& \Gamma (\bar{N}+1) \delta(t-t^\prime), \\
\braket{a_{in}^\dagger(t) a_{in}(t^\prime)} =& \Gamma \bar{N} \delta(t-t^\prime)
\end{align}
follow from the definition of $a_{in}$ as well as $\braket{a_k^\dagger(t_0) a_{k^\prime}(t_0)} = \bar{N} \delta_{k,k^\prime}$. By insertion in \eqref{eq:noiseSeparation}, these input field correlations reveal the non-vanishing noise contributions in the generalized QRT for our system.

The interferometer is arranged in such a way that one path (blue) is longer than the other (red) by a length $c T$, where $c$ is the speed of light. Signals propagating along the longer path therefore have to travel for an additional duration $T$ before reaching the second beamsplitter compared to the shorter path. We assume the beamsplitters to be lossless and have transmission (reflection) coefficients $t_i$ ($r_i$), where $i = 1$~($2$) denotes the beamsplitter closest to the emitter~(detectors).

\subsection{Detector signals \& second order correlation functions}
Using the notation of \fref{fig:photodetectionSetup}, the following relations between interferometer outputs and inputs may be observed:
\begin{align}
a_A(t) =& t_1 t_2\, a_1(t) -r_1 t_2\, a_2(t) \nonumber\\
& + r_1 r_2\, a_1(t-T) + t_1 r_2\, a_2(t-T), \label{eq:fieldRelations1}\\
\nonumber\\
a_B(t) =& -t_1 r_2\, a_1(t) + r_1 r_2\, a_2(t) \nonumber\\
&+ r_1 t_2\, a_1(t-T) + t_1 t_2 a_2(t-T),\label{eq:fieldRelations2}
\end{align}
where we have assumed that $t,r$ are real (positive) numbers and that a $\pi$-phaseshift occurs upon reflection for one of the two beamsplitter inputs (experimentally the low-to-high refraction index interface). We have ignored the effects of free-space propagation for the shared path length such that only the difference in travel time $T$ between paths is examined.

In \eqref{eq:fieldRelations1} and \eqref{eq:fieldRelations2} the consequence of adding a delayed path is immediate: the field operators at the detectors are related to field operators at the emitter (and the vacuum input) at two distinct times $t$ and $t - T$. For $T \neq 0$ the detection of a photon in either detector will therefore cause a non-trivial backaction on the state of the emitter at two different times separated by $T$.
According to Glauber's photodetection theory, the normal-ordered second order correlation function for detector A signals is on the form
\begin{equation}
G^{(2)}_A(\tau) = \braket{a_A^\dagger(t) a_A^\dagger(t+\tau) a_A(t+\tau) a_A(t)}, \label{eq:detectorAg2}
\end{equation}
where the expectation value is taken with respect to the steady-state of the system and $\tau \geq 0$. The second order correlation function describes the probability of two incident photons occurring with a temporal separation $\tau \geq 0$.

Using \eqref{eq:fieldRelations1} we may rewrite \eqref{eq:detectorAg2} in terms of $a_1$ and $a_2$ at different times. $a_2$ represents vacuum noise entering the interferometer dark port, hence we may set all terms containing $a_2$ to zero as the ordering of creation and annihilation operators is preserved by \eqref{eq:fieldRelations1}. The resulting expression for $G^{(2)}_A$ is still too cumbersome to bring here, however of the 16 constituent terms of $G^{(2)}_A$ only three are always normal-ordered, while the remaining 13 terms are for certain values of $\tau$ and $T$ out-of-time-ordered correlation functions. One example is the term
\begin{equation}
(t_1 t_2 r_1 r_2)^2\braket{a_1^\dagger(t) a_1^\dagger(t+\tau-T) a_1(t+\tau) a_1(t-T)} \label{eq:g2Term}
\end{equation}
which, depending on the relation between $\tau$ and $T$, has creation and annihilation operators appearing in different temporal orderings. For $T = 0$ all terms reduce to re-scaled versions of the correlation function $G_{a_1}^{(2)}(\tau)$ of the input field $a_1(t)$ to the interferometer.

It follows from the interaction Hamiltonian \eqref{eq:interactionHamiltonian} that $a_1(t)$ is proportional to the lowering operator $\sigma_-(t)$ of the emitter. This allows us to compute $G^{(2)}_A$ using only operators acting on the emitter state, e.g. the term \eqref{eq:g2Term} becomes proportional to the emitter operator correlation function
\begin{equation}
\braket{\sigma_+(t) \sigma_+(t+\tau-T) \sigma_-(t+\tau) \sigma_-(t-T)}.
\end{equation}
This in turn enables the use of our extended QRT in calculating $G^{(2)}_A$.

\subsection{Coherently driven emitter}
\begin{figure}
\centering
\includegraphics[scale=0.9]{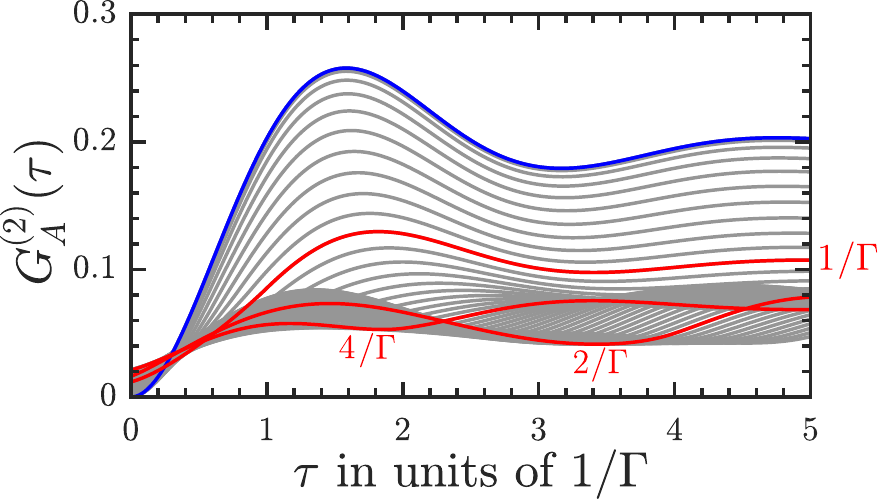}
\caption{$G^{(2)}_A$-function (\ref{eq:detectorAg2}) shown as a function of $\tau$ for different values of the interferometric delay $T$. We have used $\delta = 0$, $\Omega = 3\Gamma$, and $\bar{N} = 0$. The blue curve is $T = 0$, with red curves marking $T = 1/\Gamma$, $2/\Gamma$, and $4/\Gamma$. Grey curves show intermediate values in increments of $0.1/\Gamma$. For $T \neq 0$ the delayed path allows two photons to be detected simultaneously.}
\label{fig:detectorSignalCoherent}
\end{figure}

We first consider the case of the emitter being coherently driven by a classical field while interacting with the vacuum radiation field. \fref{fig:detectorSignalCoherent} shows $G_A^{(2)}$ for $T = 0$ (blue curve) as well as for non-zero delays $T$ in increments of $0.1/\Gamma$ (grey and red curves, with the red curves corresponding to $T = 1/\Gamma$, $2/\Gamma$, and $4/\Gamma$). We have used $\delta = 0$, $\Omega = 2\Gamma$, and $\bar{N} = 0$. The $T = 0$ (blue) curve may be recognized as the usual two-time intensity-intensity correlation function for a coherently driven two-level emitter, with the signature photon anti-bunching at $\tau = 0$.

For $T \neq 0$ we observe a non-vanishing probability of detecting two coinciding photons, as indicated by $G^{(2)}_A(0) \neq 0$. This occurs when two photons being emitted with a temporal separation~$T$ coincide at the detector after taking different paths in the interferometer.

\begin{figure}
\centering
\includegraphics[scale=0.9]{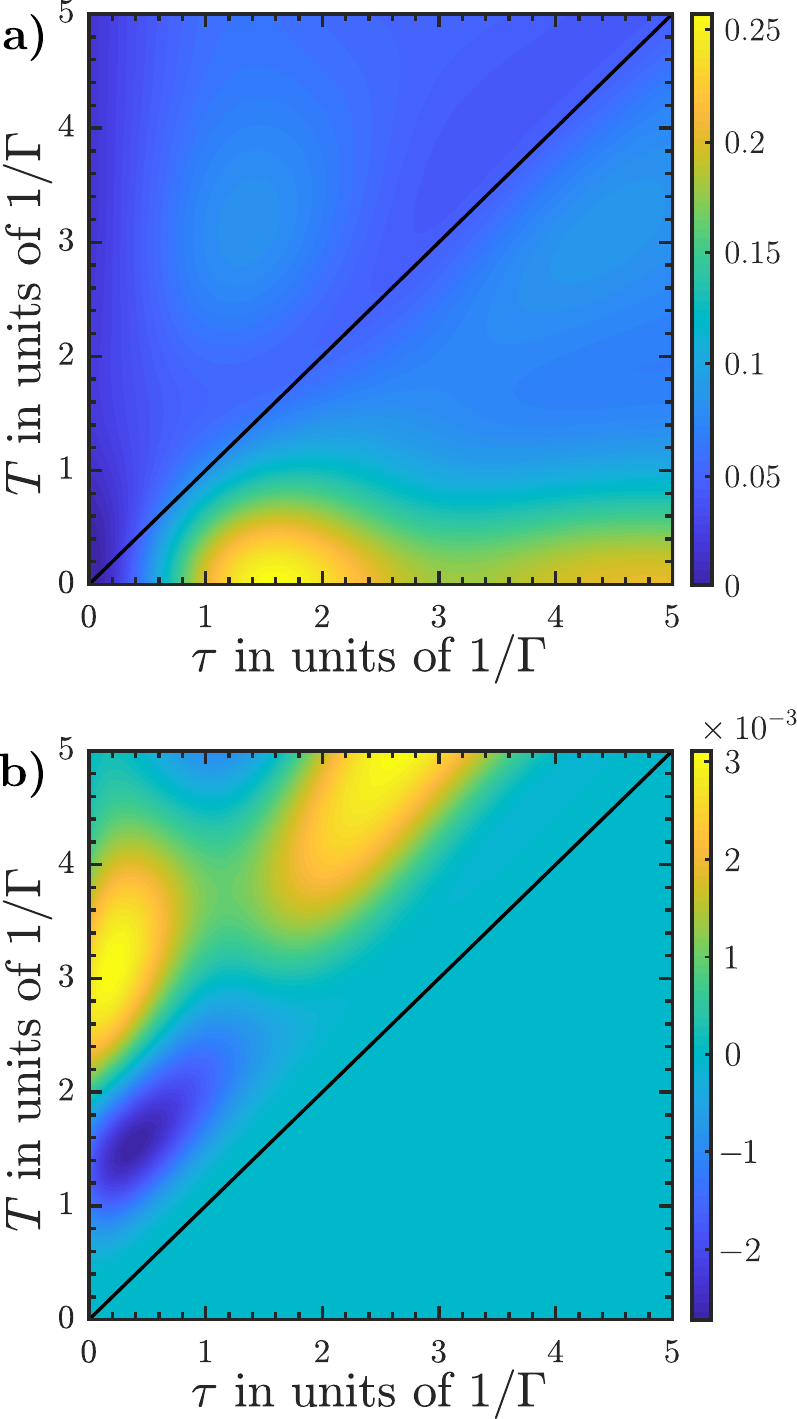}
\caption{\textbf{a):} $G^{(2)}_A$-function (\eqref{eq:detectorAg2}) as a function of detection time separation $\tau$ and delay time $T$ for a coherently driven emitter. We assume $\delta = 0$, $\Omega = 2\Gamma$, $\bar{N} = 0$. \textbf{b):} The difference $G^{(2)}_A - G^{(2)}_{A,\textbf{ no noise}}$ between calculating the $G^{(2)}_A$-function with and without inclusion of the noise terms in \eqref{eq:fourtimeCorrelationFunctionObjectEoM} (see text). In both figures the black line at $\tau = T$ separates the normal-ordered region ($\tau > T$) from the out-of-time-ordered region ($\tau < T$). Noise contributions in \eqref{eq:fourtimeCorrelationFunctionObjectEoM} only contribute in the out-of-time-ordered region, as we may observe in (b).}
\label{fig:detectorSignal2Coherent}
\end{figure}

In \fref{fig:detectorSignal2Coherent}(a) we plot $G^{(2)}_A(\tau)$ for a range of delays $T$, while \fref{fig:detectorSignal2Coherent}(b) visualizes the difference $G^{(2)}_A - G^{(2)}_{A,\text{ no noise}}$ between the true correlation function $G^{(2)}_A$ and the correlation function $G^{(2)}_{A,\text{ no noise}}$ calculated with the noise terms in \eqref{eq:fourtimeCorrelationFunctionObjectEoM} omitted. The $G^{(2)}_A$-function experiences an overall reduction of amplitude with increasing delay $T$, with the largest amplitudes occurring for small delays $T$.

In the $(\tau, T)$-plane the line $\tau = T$ (black line in both figures) separates the region of preserved time order ($\tau > T$) from the region of broken time order ($\tau < T$). In the former region the second detected photon will always have been emitted at a later time than the first one, while in the latter region the second detected photon could have been emitted prior to the first one. It is this reversed emission order that yields contributions to our measured correlations from OTOCs when we cross the transition from normal-ordering to out-of-time-ordering at $\tau = T$. In particular, as observed in \fref{fig:detectorSignal2Coherent}(b), the extra noise contributions to $G^{(2)}_A$ are present only for $\tau > T$.

\subsection{Incoherently pumped emitter}
We now consider incoherent pumping of the emitter, e.g., through interaction with a thermal reservoir ($\bar{N} > 0$, $\Omega = \delta = 0$) as illustrated in \fref{fig:photodetectionSetup}. We assume $\bar{N} = 1$ to ensure system dynamics timescales long enough to resolve OTOC particularities.

\begin{figure}
\centering
\includegraphics[scale=0.9]{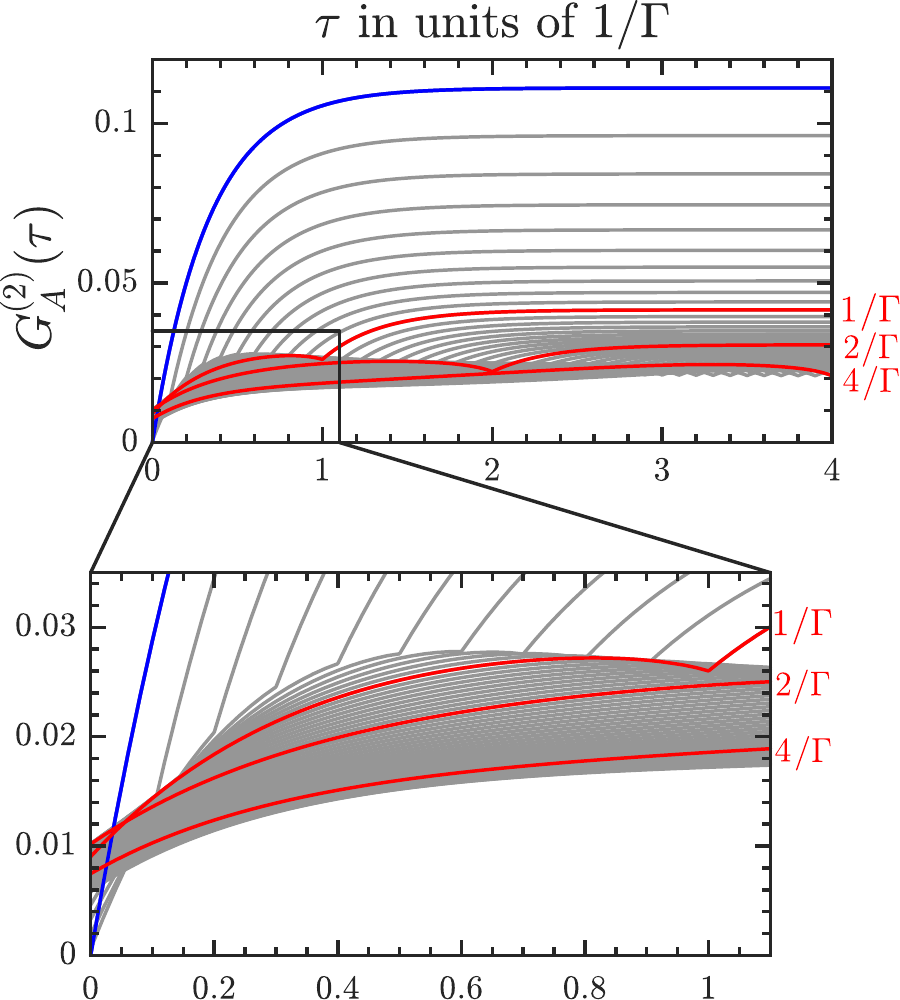}
\caption{\textbf{Main figure:} $G^{(2)}_A$-function \eqref{eq:detectorAg2} shown as a function of detection time separation $\tau$ for different values of the delay $T$ from $0$ to $4/\Gamma$ in increments of $0.1/\Gamma$. We are considering the case of incoherent pumping, with $\bar{N} = 1$, $\Omega = \delta = 0$. The blue curve is $T = 0$, and red curves mark $T = 1/\Gamma$, $2/\Gamma$, and $4/\Gamma$. Grey curves correspond to intermediate values of $T$ in increments of $0.1/\Gamma$. As in \fref{fig:detectorSignalCoherent} the anti-bunched nature of the emitter signal is lifted for non-vanishing delays $T$. \textbf{Zoomed figure:} We observe a cusp for each of the grey and red curves, corresponding to the point $\tau = T$. These cusps are caused by OTOC constituents of $G^{(2)}_A$ which vanish for $\tau < T$.}
\label{fig:detectorSignalThermal}
\end{figure}

In \fref{fig:detectorSignalThermal} we show the $G_A^{(2)}$-function for $T = 0$ (blue, upper curve) as well as for non-zero delays $T$ in increments of $0.1/\Gamma$ (grey and red curves, with the red curves corresponding to $T = 1/ \Gamma$, $2/\Gamma$, and $4/\Gamma$). As in the coherent case (\fref{fig:detectorSignalCoherent}), the zero delay $G^{(2)}_A$-function is the usual two-time intensity-intensity correlation function of a now incoherently pumped two-level emitter. For increasing delays $T > 0$, the photon anti-bunching is lifted as in the case of the coherently driven emitter.

The general shape of the correlation function is distorted with increasing $T$, and one prominent feature of the $G^{(2)}_A$-function is the cusp visible for $\tau = T$ for all $T \neq 0$ curves. As discussed previously, the point $\tau = T$ marks a transition from OTOCs to normal-ordered correlation functions. In this particular case, several of the out-of-time-ordered $G^{(2)}_A$ constituents are zero for $\tau < T$ and non-zero for $\tau > T$, hence causing the cusp in the $G^{(2)}_A$-function.

\begin{figure}
\centering
\includegraphics[scale=0.9]{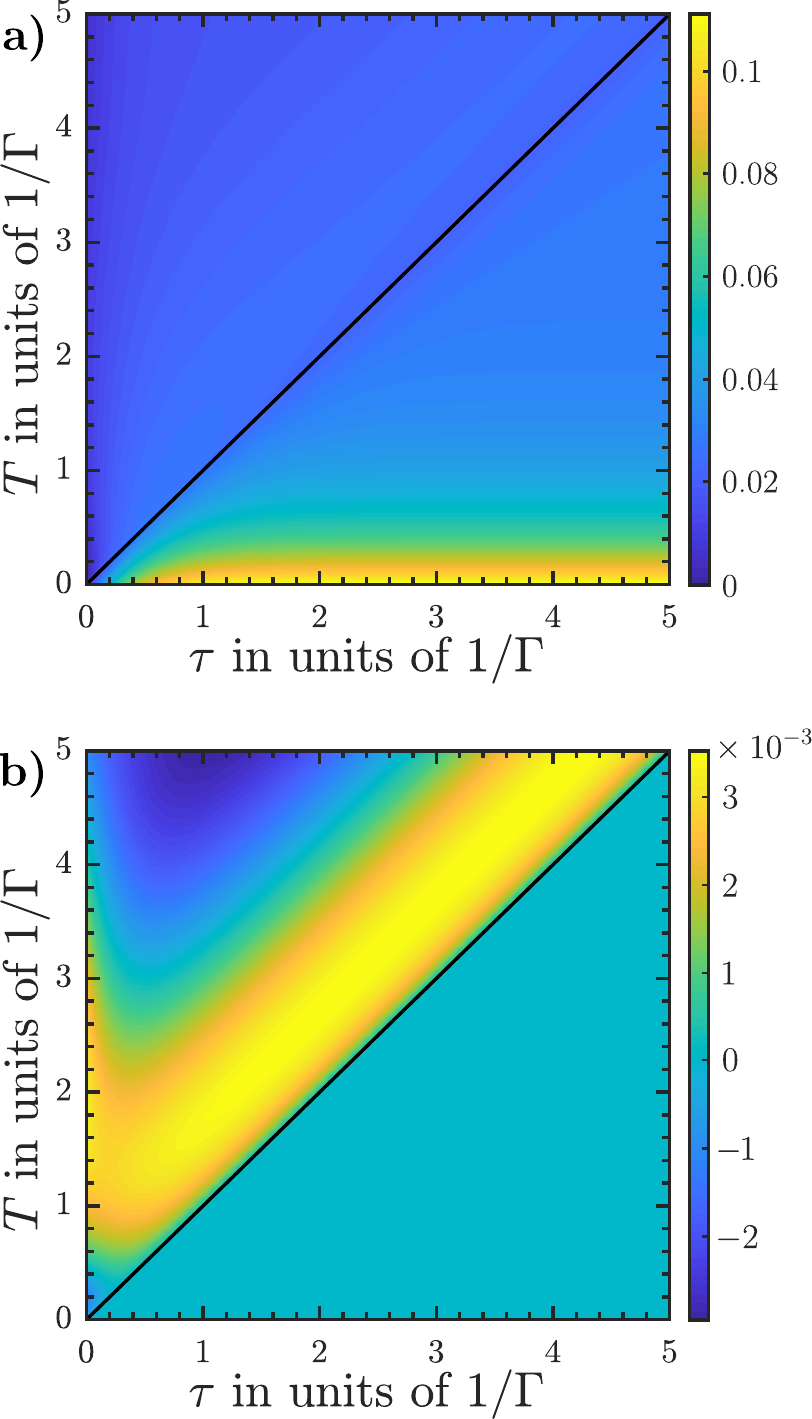}
\caption{\textbf{a):} $G^{(2)}_A$-function \eqref{eq:detectorAg2} as a function of detection time separation $\tau$ and delay time $T$ for an incoherently pumped emitter ($\bar{N} = 1$). \textbf{b):} The difference $G^{(2)}_A - G^{(2)}_{A,\text{ no noise}}$ between $G^{(2)}_A$ calculated with and without noise terms in \eqref{eq:fourtimeCorrelationFunctionObjectEoM}. The black line in both figures separates the normal-ordered region $\tau > T$ from the out-of-time-ordered region ($\tau < T$). In accordance with the discussion in section~\ref{section:application}, we see the noise contributions to the QRT calculation only for $\tau < T$.}
\label{fig:detectorSignal2Thermal}
\end{figure}

In \fref{fig:detectorSignal2Thermal}(a) the $G^{(2)}_A$-function is plotted for a range of delays $T$, and we again observe an overall reduction of amplitude with increasing delay $T$, with the largest amplitudes of $G^{(2)}_A$ occurring for small delays $T$ and large detection separations $\tau$. In \fref{fig:detectorSignal2Thermal}(b) we plot the difference $G^{(2)}_A - G^{(2)}_{A,\text{ no noise}}$ between the $G^{(2)}_A$-function calculated with and without noise terms in \eqref{eq:fourtimeCorrelationFunctionObjectEoM}. As in \fref{fig:detectorSignalThermal}(b) we clearly see the difference between regions $\tau > T$ and $\tau < T$, with notable contributions from the extra OTOC noise terms in the latter region.

\section{Discussion}\label{section:discussion}
The quantum regression theorem (QRT) is a powerful tool for the calculation of correlation functions for Markovian open quantum systems. While previously the QRT has been restricted to normal-ordered correlation functions we have in this article demonstrated the extension of the QRT to the calculation of out-of-time-ordered correlation functions.

This extension allows us to describe the effect of induced delays within optical detection, as demonstrated in section~\ref{section:application} for a Mach-Zender-like detection scheme. As also demonstrated in Schrama \textit{et al.} \cite{Schrama1992} and Bali \textit{et al.} \cite{Bali1993}, our results show that the normal-ordered correlation functions for the detector signal depend on out-of-time-ordered correlation functions (OTOCs) for the emitting system when the signal experiences  non-uniform travel delays.

With the recent interest in OTOCs and their ability to capture system properties and behavior, e.g. in quantifying the scrambling of information in a system, it is necessary from a theoretical point of view to be able to calculate OTOCs for experimentally relevant setups. Though prior investigations have calculated particular OTOCs for open quantum systems\cite{Schrama1992,Bali1993,Syzranov2018}, the extended QRT derived in this article provides a unified procedure for calculating arbitrary OTOCs in any Markovian open quantum system.

While some experimental systems may allow an effective time reversal through modification of the sign of the system Hamiltonian, open quantum systems do not in general admit time reversal due to the interaction with their environment and hence cannot extract OTOCs by Loschmidt echo techniques. Instead we believe that our interferometric technique to impose delays may inspire approaches to identify OTOC contributions to signals measured in a normal-ordered sense. Combining the signals from both detectors in section~\ref{section:application} by e.g. subtracting the signal of detector B from detector A, we arrive at a correlation function $G^{(2)}_{A-B}(\tau)$ which for $\tau < T$ is a sum of solely out-of-time-ordered terms. By varying the transmittance and reflectance of the two beamsplitters, we can further manipulate this $G^{(2)}_{A-B}$-function to extract combinations of only a few OTOCs.

\section{Acknowledgments}
The authors acknowledge financial support from the Villum Foundation.

\end{document}